	 \documentclass[12pt,preprint]{aastex}

	\newcommand{\kms}{km\thinspace s$^{-1}$}

\received{Dec. 2002} \accepted{** 2003} \journalid{337}{15 January 1989}
\articleid{11}{14} \slugcomment{}

\begin{document}

\title{Multiple Molecular H$_2$ Outflows in AFGL~618\altaffilmark{1} }

\author{
	P. Cox\altaffilmark{2}, 
	P.J. Huggins\altaffilmark{3},
	J.-P. Maillard\altaffilmark{4}, 
	C. Muthu\altaffilmark{5},		
	R. Bachiller\altaffilmark{5},
	T. Forveille\altaffilmark{6,7}
       }
 
\altaffiltext{1}{Based on observations collected at the
        Canada-France-Hawaii Telescope, operated by the National
        Research Council of Canada, the Centre National de la
        Recherche Scientifique de France, and the University of
        Hawaii.}
\altaffiltext{2}{Institut d'Astrophysique Spatiale, 
  Universit\'e de Paris Sud, F-91405 Orsay, France \ Pierre.Cox@ias.u-psud.fr}
\altaffiltext{3}{Physics Department, New York University, 4 Washington Place,
	New York, NY 10003 \ patrick.huggins@nyu.edu }  
\altaffiltext{4}{Institut d'Astrophysique de Paris, C.N.R.S., 
	98b bd. Arago, F-75014 Paris, France \ maillard@iap.fr}
\altaffiltext{5}{Observatorio Astron\'omico Nacional (IGN), 
	Apt 1143, 28800 Alcal\'a de Henares, Spain \ muthu@oan.es bachiller@oan.es}
\altaffiltext{6}{CFHT, PO Box 1597, Kamuela, HI 96743, USA \ forveill@cfht.hawaii.edu}
\altaffiltext{7}{Observatoire de Grenoble, B.P. 53X, 
	38041 Grenoble Cedex, France}

\begin{abstract}

We report high spatial (0\farcs5) and high spectral (9~\kms) resolution
spectro-imaging of the 2.12~$\mu$m H$_2$ 1$\rightarrow$0~S(1) line in
the proto-planetary nebula AFGL~618 using \emph{BEAR} at the CFHT. The
observations reveal the presence of multiple, high-velocity, molecular
outflows that align with the remarkable optical jets seen in
\emph{HST} images.  The structure and kinematics of the outflows show
how jets interact with circumstellar gas and shape the environment in
which planetary nebulae form.

\end{abstract}


\section{Introduction}

High velocity ($\gtrsim 100$~\kms) directed outflows play a dramatic
role in the rapid evolution of stars from the asymptotic giant branch
(AGB) to the planetary nebula (PN) phase. The outflows interact with
the roughly spherical molecular envelopes ejected by the stars on the
AGB, producing strong shocks that pierce and disrupt the circumstellar
gas (e.g., Cox et al. 2000; Huggins et al. 2000).  Observations of
shock diagnostics in objects that have only recently left the AGB are
therefore of considerable interest in constraining this crucial
evolutionary phase.

AFGL~618 is a bipolar proto-planetary nebula (PPN), which evolved from
the AGB about $\rm 200 \, yr$ ago (Kwok \& Bignell 1984). It was first
identified by Westbrook et al. (1975) and since then it has been
extensively observed. The presence of high-velocity molecular gas has
been reported extending over $\sim$250~\kms\ in H$_2$
{1\,$\rightarrow$\,0 S(1)} (Burton \& Geballe 1986; Kastner et
al. 2001) and $\sim$400~\kms\ in CO (Cernicharo et al. 1989; Gammie
et al. 1989). High velocities are also seen in optical lines (e.g.,
Carsenty \& Solf 1982) and the outflows appear as spectacular jet
features in the \emph{HST} images published by Trammell \& Goodrich
(2002). In this \emph{Letter}, we report high velocity resolution
($\rm 9 \, km \, s^{-1}$) spectro-imaging of AFGL~618 in the H$_2$
{1\,$\rightarrow$\,0 S(1)} line that reveals the detailed structure
and kinematics of the shocked molecular gas, and its relation to the
optical jets.

\section{Observations}

The observations were made on October 6, 2000 using the \emph{BEAR}
Imaging FTS at the f/35 infrared focus of the 3.60-meter
Canada-France-Hawaii Telescope.  A detailed description of \emph{BEAR}
and the data reduction is given by Maillard (2000). The 256$\times$256
HgCdTe facility camera provides a circular field of view of
24$^{\prime\prime}$ diameter, and the plate scale on the detector is
0\farcs35/pixel.  The seeing at 2~$\rm \mu m$ during the observations
was typically 0\farcs5.  AFGL~618 was observed through a narrow band
filter ($\rm 4700 - 4750 \, cm^{-1}$) which includes the H$_2$
{1\,$\rightarrow$\,0 S(1)} line at 4712.9~cm$^{-1}$ (2.12~$\rm \mu
m$).  The raw data consist of a cube of 1150 planes with an
integration time of 9~sec per image, an image being taken at each step
of the interferometer.  The maximum path difference corresponds to a
resolution (FWHM) of 0.137~cm$^{-1}$, i.e. 8.9~km~s$^{-1}$.  A data
cube of the star HD~162208 ($\rm m_K$~=~7.11) was obtained for
calibration. The 2~$\rm \mu m$ continuum emission of AFGL~618, which
is found to be strongly peaked toward the center, has been subtracted
from each plane of the cube.  We have used the \emph{HST} images and the
J-band and H$_2$ images of Ueta et al. (2001) to determine the
co-ordinates of the \emph{BEAR} data.

\section{The Molecular Outflows}

\subsection{Overview of the H$_2$ Emission}

The velocity integrated image of AFGL~618 in the H$_2$
{1\,$\rightarrow$\,0 S(1)} line is shown in Fig.~1, together with the
\emph{HST} image in H$\alpha$.  The H$_2$ emission is dominated by two
bipolar lobes which are oriented at $\rm PA \sim 104\arcdeg$, and
correspond to the jet system seen in the H$\alpha$ emission of
the shocked-ionized gas. 

The kinematic structure of the H$_2$ emission is found to be complex,
and we present different aspects of the data in Figs.~2--5.  The top
panel in Fig.~2 shows the H$_2$ {1\,$\rightarrow$\,0 S(1)} spectrum of
the whole nebula. The emission peaks at a velocity of $-34.5 \pm
0.4$~\kms\ (LSR), and shows broad wings extending from $-200$ to
+160~\kms, in agreement with previous observations (e.g., Burton \& Geballe
1986).  The peak is blue-shifted relative to the systemic velocity
measured in CO ($-21.8$~\kms, Cernicharo et al. 1989), probably
because of extinction.  At low expansion velocities, maps of the H$_2$
are roughly symmetric about the center, but at higher velocities there
is a clear blue/red separation to the east and west, respectively
(Fig.~3). The nebula in H$_2$ is therefore tilted to the line of sight
toward us in the east, consistent with the optical appearance.

The strongest H$_2$ emission lies $\sim 2\arcsec$ each side of the
center near the systemic velocity, but the emission near these
positions also extends to the highest velocities. This is seen in
spectra at these two positions (Fig.~2, center), which also show steep
sided profiles in the red and blue wings to the east and west,
respectively.  The relation of these spectra to the overall
kinematics is seen in the position-velocity diagram along the nebula
axis shown in Fig.~4 -- see also Kastner et al. (2001).

\subsection{Multiple Molecular Outflows} 

In addition to the main kinematic structure described above, a
striking aspect of our data is the presence of localized,
high-velocity features in the $\rm H_2$ emission which are associated
with the optical jets seen in the H$\alpha$ image.  Fig.~1 already
suggests some relation between the $\rm H_2$ and H$\alpha$ emission,
and this becomes evident in the $\rm H_2$ channel maps at the extreme
red- and blue-shifted velocities (Fig.~3). At these high velocities,
discrete structures trace the locations of individual optical jets
seen in the underlying \emph{HST} image. These features can be
followed in the data cube in position and velocity, and delineate the
interaction of the jets with the molecular envelope.

On account of the tilt of AFGL~618 to the line of sight and the high
optical extinction toward the equatorial regions ($A_V \, \gtrsim \,
10 \, \rm mag.$, e.g., S\'anchez Contreras et al. 2002), there is less
obscuration of the east lobe of the nebula. The optical jets are best
seen on this side, and it is here that they show the clearest
correspondence with the high velocity features seen in $\rm H_2$. The
correspondence is less complete in the west lobe, probably in part due
to the higher optical extinction near the center, but also as a result
of real differences (e.g., high velocity $\rm H_2$ lies part way along
jet {\em a} in Fig.~3, but not along the whole length) which could
reflect the geometry and/or strength of the interactions.

Fig.~5 presents a sequence of channel maps that show the relation
between the optical jets and the H$_2$ in the east lobe in detail.
The origin of the optical jets cannot be traced near the star due to
extinction, but $\rm H_2$ at intermediate velocities ($-100$~\kms) is
detected near the base of the jets where they diverge from the center.
At successively higher velocities, the $\rm H_2$ develops into
separate, localized, structures that follow individual optical
jets. Four separate structures can be seen corresponding to all four
jets identified by Trammel \& Goodrich (2002) in the east lobe ({\emph
b}, {\emph c}, {\emph d}, and {\emph e}).  Jet {\emph b} is traced out
to $-163$~\kms. Jets {\emph c} and {\emph e} appear to be connected,
with jet {\emph c} terminating at $-$163~\kms, and the associated jet
{\emph e} with an additional component extending to $-178$~\kms. The
highest velocities ($-200$~\kms) are associated with jet {\emph d},
which is short and protrudes near the base of the jet system.

Three features of the observations provide further information on the
relation of the jets to the molecular emission. First, although the
$\rm H_2$ emission is not well spatially resolved across the jets, it
is resolved along the extended jets {\emph b} and {\emph c}+{\emph e},
and shows a velocity gradient, increasing away from the
center. Second, unlike the $\rm H\alpha$, the $\rm H_2$ emission is not
strongly peaked at the head of the optical jets but extends downstream
toward the star. Third, the $\rm H_2$ line widths observed near the
heads of the jets are narrow ($\approx 30$~\kms) compared with the
bulk velocity of the gas relative to the systemic velocity (Fig.~2,
bottom panel).

\section{Discussion}

\subsection{Geometry of the Outflows}

The overall distribution and kinematics of the $\rm H_2$ suggests that
the main emission arises in roughly formed, biconical cavities,
consistent with the overall picture inferred from optical images. The
steep-sided spectra (Fig.~2) and the sudden change from red- to
blue-shifted gas near the systemic velocity (Fig.~3 \& 4) indicate
that the far side of the approaching (east) lobe and the near side of
the receding (west) lobe lie close to the plane of the sky. The
highest velocity gas then arises near the center, in the sides of the
lobes that lie closest to the line of sight.

The east-west (blue-red) tilt of the optical jets (e.g., S\'anchez
Contreras et al. 2002) which form the lobes, and the angles of
individual jets with respect to the line of sight estimated by Trammel
\& Goodrich (2002) (51\arcdeg, 31\arcdeg, and 54\arcdeg\ for jets
{\emph b}, {\emph c}, and {\emph d}, respectively) are consistent with
this general picture, although the short projected length of jet
{\emph d}, and the very high velocity that we observe in $\rm H_2$
suggest that it may be the closest to the line of sight. Given the
short lifetime for optical emission from the jets (a few years, e.g.,
S\'anchez-Contreras et al. 2002) compared to their dynamical time
scales of $\lesssim 200$~yr (see \S~4.2), there could also be a
history of outflows which partially excavated the biconical cavities
in addition to those currently seen in the \emph{HST} image.

\subsection{Jet Interactions}

Our observations cast light on the long-standing problem of the origin
of the very high velocity H$_2$ emission in AFGL~618 (e.g., Burton \&
Geballe 1986; Hartquist \& Dyson 1987). Figs.~3 and 5 show that the
emission arises from the interaction of jets with the circumstellar
gas.  The presence of $\rm H_2$ emission along the length of the
optical jet structures is consistent with a class of models in which
the ambient gas is shocked and dragged along with the jet through
prompt or steady entrainment (e.g., Hartigan et al. 1996); similar
cases are seen in the outflows around young stellar objects (e.g.,
Bachiller 1996). The kinematics that we observe place valuable
constraints on such a picture. The $\rm H_2$ line widths near the
heads of the jets ($\approx 30$~\kms) set upper limits on the
turbulence which is believed to play a key role in the entrainment
process. On the other hand, the bulk velocity of the molecular gas is
extremely high. For example, the observed radial velocity of the $\rm
H_2$ just behind the head of jet {\emph b} is $\sim 140$~\kms\
(relative to the systemic velocity); if the inclination to the line of
sight is $51\arcdeg$ (see above), the actual velocity is $\sim
220$~\kms, and if the angle of inclination is $66\arcdeg$ (S\'anchez
Contreras et al. 2002) the velocity is 340~\kms. For an adopted
distance of 1.5~kpc to AFGL~618, the dynamical timescale of the jet is
$\lesssim 200$~yr.

The velocities are much too large for the molecular gas to survive
acceleration in a single shock, for which the maximum velocity is
$\sim 45$~\kms\ (Draine, Roberge, \& Dalgarno 1983).  However, it is
possible for the molecules to reform in the post-shock gas (e.g.,
Hollenbach \& McKee 1989; Neufeld \& Dalgarno 1989) although models of
high velocity bow shocks including molecule reformation have not yet
been developed. 
 
It is striking that the H$_2$ velocities that we observe are
significantly higher than those indicated by observations of optical
lines. The interpretation of optical line ratios at the heads of the
jets using shock models yields shock velocities of only 40--100~\kms\
(Trammell \& Goodrich 2002).  The observed radial velocities in
optical lines at the head of jet {\emph b} are $\sim 80$~\kms\
(S\'anchez Contreras et al. 2002), and when corrected for an
inclination of 51\arcdeg--66\arcdeg\ (see above) give values of $\sim
130$--200~\kms, the latter similar to shock velocities estimated from
the optical line widths (S\'anchez Contreras et al. 2002).  However,
independent of the inclination angle, the observed $\rm H_2$ velocity
in jet {\emph b} is a factor of $\sim 1.8$ higher than the optical
lines at the head.

If the H$_2$ emission that we observe is shocked excited circumstellar
gas, as seems most likely, then the high velocities that we observe
suggest an even higher velocity, collimated driving wind. In addition,
the differences between the H$_2$ and optical line velocities indicate
a range in shock conditions, which is not unlikely given the small
scale structure seen in the H$\alpha$ image.  A possible, alternative
scenario for the origin of the high velocity molecular gas in one in
which the molecules are entrained or formed in the outflows close to
their source, so are already present at high velocity and are excited
by interactions with the ambient gas in the regions observed.
Spectro-imaging of the optical lines with both higher spectral and
spatial resolution than are currently available will be important to
further study these issues.

\subsection{Evolution of the Nebula}

The H$_2$ observations reported here trace the detailed structure and
kinematics of the molecular outflows in AFGL~618 which are also seen
at lower angular resolution in other probes of the molecular gas.  The
full H$_2$ spectrum shown in Fig.~2 resembles that of the
CO lines (e.g., Cernicharo et al. 1989), and the shift from blue to red in
going from the east to the west side of the nebula (Fig.~3) is also
seen in millimeter molecular maps (e.g., Neri et al. 1992; Hajian et
al. 1996; Meixner et al. 1998).  However, the CO bullets reported by
Ueta et al. (2001) at high negative velocities on the west side of the
nebula are at variance with the overall kinematics reported here, and
are probably not real.

The present observations provide evidence that the high velocity H$_2$
outflows in AFGL~618 are the result of the interaction of jets with
the circumstellar envelope.  Similar examples of this type of
interaction at different stages of development are seen in
AFGL~2688 (Cox et al. 2000), M1-16 (Huggins et al. 2000), and NGC~7027
(Cox et al.  2002).  All these cases support a view in which the
geometry of the circumstellar gas around a newly forming PN is largely
determined by the action of the jets, and plays a key role in
determining the morphology of the fully formed PN.

\acknowledgements 
We thank S. Trammell for providing a copy of the \emph{HST} H$\alpha$
image. We also thank A. Abergel, T. Ueta, and G. Pineau des For\^ets
for useful discussions, and an anonymous referee for helpful
comments. This work has been supported in part by NSF grant
AST-9986159 and a visiting professorship at the Universit\'e de
Paris-Sud (to P.J.H.).

\clearpage 


 \figcaption[]{ Velocity integrated H$_2$ {1\,$\rightarrow$\,0 S(1)}
image of AFGL~618 observed with \emph{BEAR} (bottom) and the
\emph{HST} image in H$\alpha$ (top) from Trammell \& Goodrich (2002).
The H$_2$ image is continuum subtracted, and the peak intensity is
$\rm 1.4 \times 10^{-5}$~W\,m$^{-2}$\,sr$^{-1}$.  The coordinates of
the field center are $\rm \alpha =04^h42^m53\fs58$, $\rm \delta =
36^{\rm o} 06\arcmin 53\farcs6$ (J2000.0). }

\vspace{0.5cm}


\figcaption[]{H$_2$ {1\,$\rightarrow$\,0 S(1)} spectra of AFGL~618.
\emph{Top}: integrated over the whole nebula.  \emph{Middle}:
2\arcsec\ east and 2\arcsec\ west (filled spectrum) of the
center.  \emph{Bottom}: near tip of jet {\emph b}. In the middle and
bottom panels, the flux density is per pixel. }

\vspace{0.5cm}


\figcaption[]{H$_2$ channel maps of AFGL~618 superposed on the
\emph{HST} H$\alpha$ image.  \emph{Top}: emission around the systemic
velocity ($-40$ to $-3$ \kms).  \emph{Middle}: blue and red
intermediate velocities ($-41$ to $-125$ and $-3$ to 81~\kms).
\emph{Bottom}: blue and red extreme velocities ($-126$ to $-186$ and
82 to 152~\kms). In the middle panel, the jets are labeled as in
Trammell \& Goodrich (2002). The contours are at 2, 6, and 10 to 100
by 20 percent of the peak intensity in each channel.}

\vspace{0.5cm}

         
\figcaption[]{Position-velocity H$_2$ map of AFGL~618,
oriented at PA = 94\fdg5 (joining jets a and b, see Fig.~2).   
The contours are at 3.5, 7, 10.5, 14, and 20 to 100 by 10  
percent of the peak intensity ($\rm 10^{-5} \, W \, m^{-2} \, sr^{-1}$) 
of the map. }
           
\vspace{0.5cm}

\figcaption[]{Sequence of H$_2$ channel maps of the east
lobe of AFGL~618 superposed on \emph{HST} H$\alpha$ image. The
channel width is 15~\kms\ and  the central velocity is given in
each panel. The contours are 1, 1.5, 2, 3 to 30 by 2.5~$\rm
\times 10^{-7}$~W\,m$^{-2}$\,sr$^{-1}$. The jets are labeled
as in Fig.~3. }

\newpage

\begin{figure}
\epsscale{0.7}
\plotone{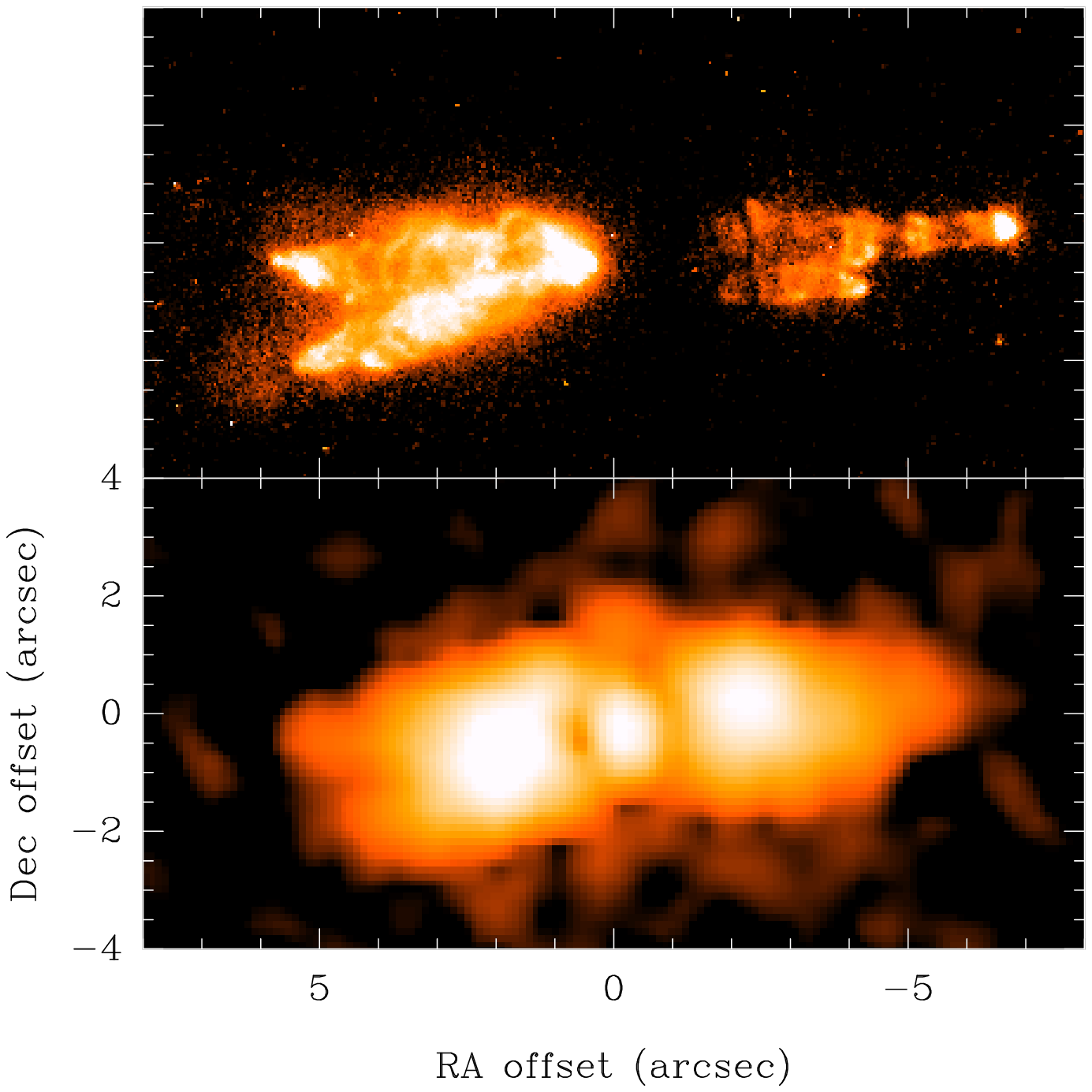}
\end{figure}
 
\clearpage
 
\begin{figure}
\epsscale{0.7}
\plotone{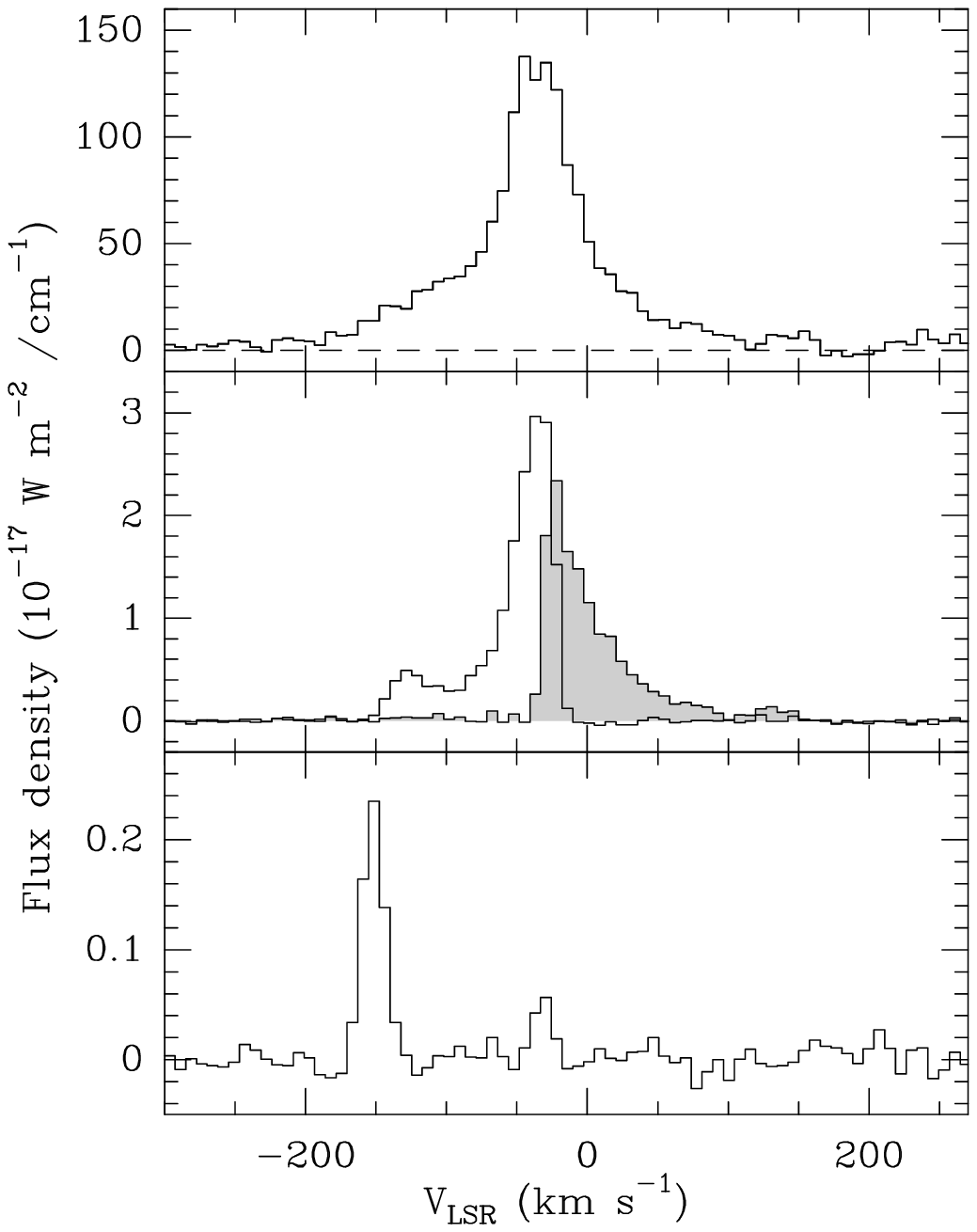}
\end{figure}

\clearpage
 
\begin{figure}
\epsscale{0.7}
\plotone{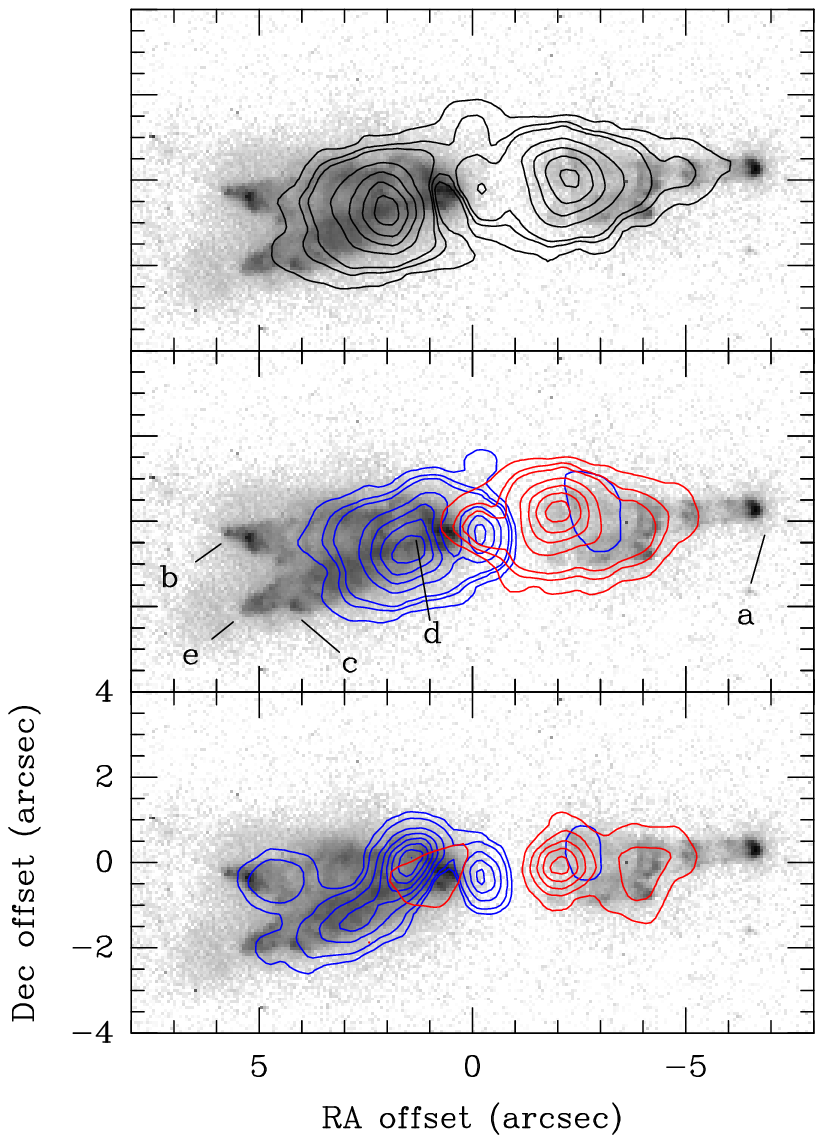}
\end{figure}

\clearpage
 
\begin{figure}
\epsscale{0.7}
\plotone{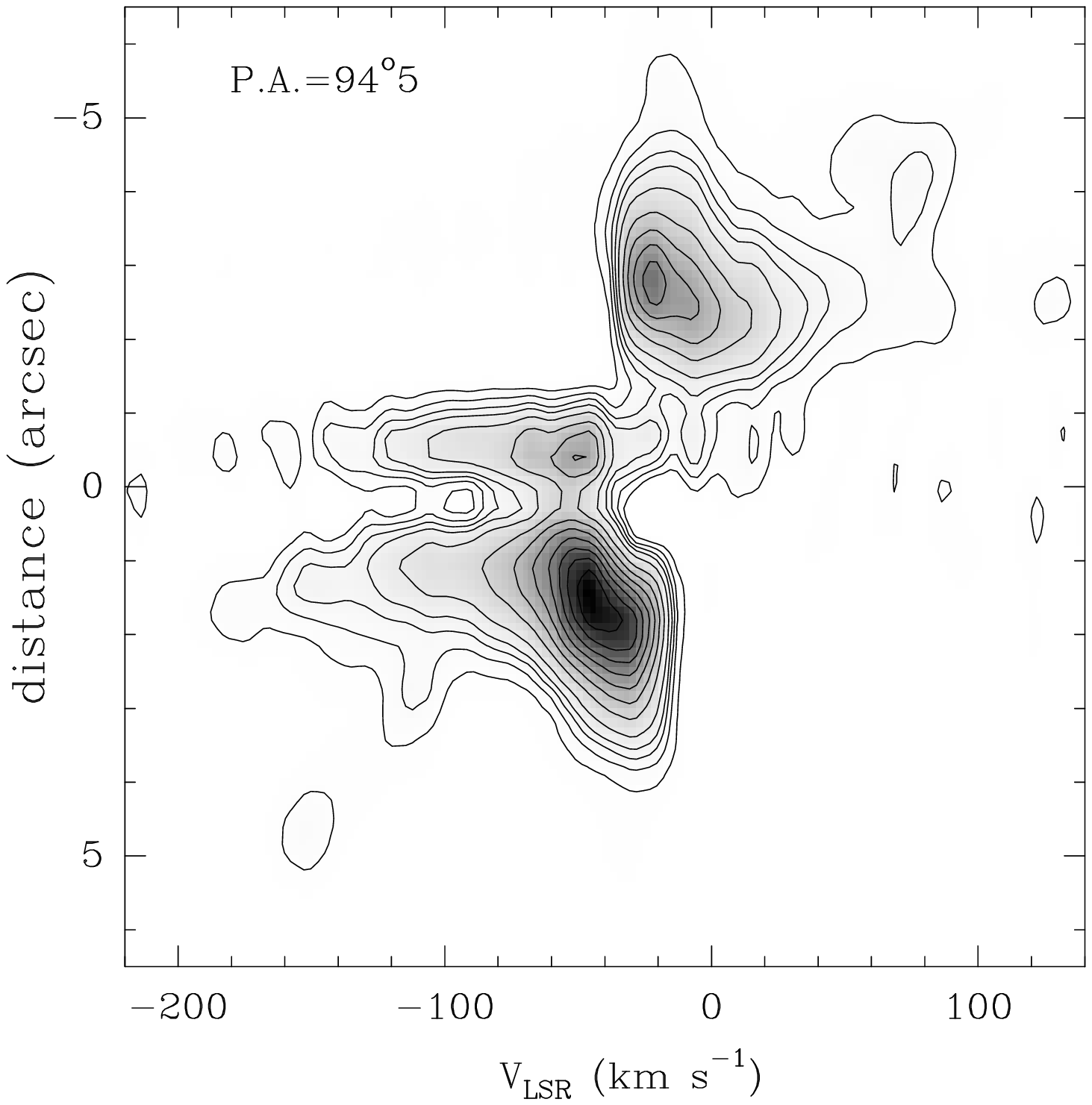}
\end{figure}

\clearpage
 
\begin{figure}
\epsscale{1.0}
\plotone{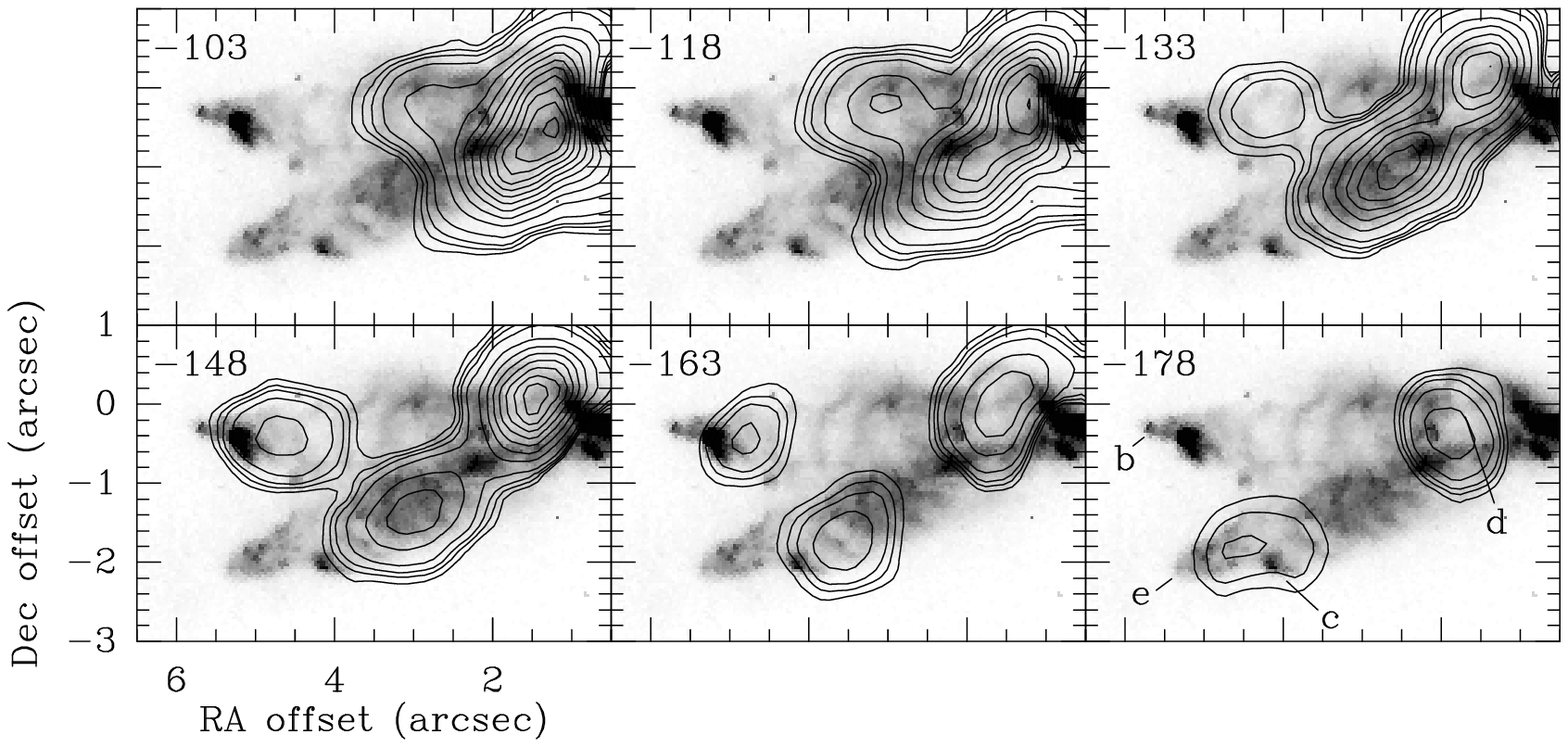}
\end{figure}

\end{document}